\documentclass[preprint,aps,showpacs,nofootinbib,superscriptaddress,preprintnumbers,epsf,psf]{revtex4}
\usepackage[english]{babel}
\usepackage{pstricks}
\usepackage[]{graphicx}
\usepackage{epsf}
\usepackage{amsmath}
\usepackage{amsfonts}
\usepackage{amssymb}

\usepackage{color}
\usepackage{dcolumn}
\usepackage{bm}
\everymath{\displaystyle}
\begin{document}

\title{Hyperon polarization in Heavy-Ion Collisions and gravity-related anomaly}

\author{\firstname{Mircea}~\surname{Baznat}}
\email{baznat@theor.jinr.ru}\affiliation{Joint Institute for Nuclear
Research, 141980 Dubna (Moscow region), Russia}
\affiliation{Institute of Applied Physics, Academy of Sciences of
Moldova, MD-2028 Kishinev, Moldova }

\author{\firstname{Konstantin}~\surname{Gudima}}
\email{gudima@cc.acad.md}
\affiliation{Institute of Applied Physics, Academy of Sciences of
Moldova, MD-2028 Kishinev, Moldova }
\author{\firstname{Alexander}~\surname{Sorin}}
\email{sorin@theor.jinr.ru}
\author{\firstname{Oleg}~\surname{Teryaev}}
\email{teryaev@theor.jinr.ru} \affiliation{Joint Institute for
Nuclear Research, 141980 Dubna (Moscow region), Russia}
\affiliation{National Research Nuclear University MEPhI (Moscow
Engineering Physics Institute), Kashirskoe Shosse 31, 115409 Moscow,
Russia} \affiliation{Dubna International University, 141980, Dubna,
Russia}
\date{\today}

\begin {abstract}
We study the energy dependence of global polarization of $\Lambda$
hyperons in peripheral $Au-Au$ collisions. We combine the
calculation of vorticity and strange chemical potential in the
framework of kinetic Quark-Gluon String Model with the anomalous
mechanism related to axial vortical effect. We pay special attention
to the temperature dependent contribution related to gravitational
anomaly and found that the preliminary RHIC data are compatible with
its suppression discovered earlier in lattice calculations.

\end{abstract}

\pacs {25.75.-q}

\maketitle

\section{Introduction}

The experimental evidences for polarization of hyperons in heavy-ion
collisions found by STAR collaboration \cite{Lisa} attracted
recently much attention
\cite{Becattini:2016gvu,Karpenko:2016jyx,Karpenko:2016idy,Xie:2015xpa,Fang:2016uds}.

The studies of polarization are often performed
\cite{Becattini:2013vja} in the framework of approach exploring
local equilibrium thermodynamics \cite{Becattini:2014yxa} and
hydrodynamical calculations of vorticity
\cite{Betz:2007kg,Csernai:2013bqa,Csernai:2014hva}.

There is another (although related \cite{Prokhorov:2017atp}) approach to polarization first proposed in
\cite{Rogachevsky:2010ys} and independently in \cite{Gao:2012ix}.
The so-called axial vortical effect (see e.g.
\cite{Kalaydzhyan:2014bfa} and references therein) being the
macroscopic manifestation of axial anomaly \cite{Son:2009tf} leads to
induced axial current of strange quarks which may be converted to
polarization of $\Lambda$-hyperons
\cite{Rogachevsky:2010ys,Gao:2012ix}.

The effect is proportional to vorticity and helicity of the strong
interacting medium, and, in particular, to helicity separation effect  discovered  \cite{Baznat:2013zx} in the
kinetic Quark-Gluon-String
Model(QGSM)\cite{toneev83,toneev90,amelin91} and confirmed \cite{Teryaev:2015gxa}  in HSD \cite{Cassing:1999es} model. 
This helicity separation effect receives  \cite{Baznat:2013zx,Teryaev:2015gxa} 
the significant contribution $ \sim \vec v_y \vec \omega_y$ from the transverse component of velocity and vorticity. It is easily explained \cite{Baznat:2013zx} by the same signs of transverse vorticities in the 
"upper" and "lower" (w.r.t. reaction plane) half-spaces, combined with the opposite signs of velocities. 
 At the same time, even larger contribution  \cite{Baznat:2013zx,Teryaev:2015gxa} of longitudinal components of velocity and vorticity $\sim \vec v_z \vec \omega_z$ implies the appearance of the
"quadrupole" structure of longitudinal vorticity, recently found \cite{Becattini:2017gcx} in the hydrodynamical approach.

Indeed, the opposite values of longitudinal velocities in the "left" and "right" (w.r.t. to "vertical" plane $x=0$ normal to reaction one and containing the beams direction) require exactly the quadrupole structure of longitudinal vorticities in the quater-spaces formed by the intersection of reaction and vertical planes: 
\begin{eqnarray}
h =h_x+h_y+h_z \sim sign(y); \\
v_z \sim sign (x); \\ 
\omega_z \sim sign(x) sign (y),  
\end{eqnarray}
where $h_i =  v_i \omega_i$ is the contribution of the respective component of velocity and vorticity to the helicity density.   
It is this quadrupole structure of vorticity that leads to the up-down mirror structure of helicity  after multiplication by the left-right mirror structure of velocity: 
$$h_z = \omega_z v_z \sim (sign(x))^2 sign(y) = sign(y).$$ 

To make (2) applicable and observe the quadrupole picture one needs to average the longitudinal velocity and vorticity over the cylinder symmetric
w.r.t. the plane $z=0$. Otherwize, taking the different slices $z=const$, longitudinal velocity is not, generally speaking,  changing sign with $x$.
The dependence of the quadrupole picture over the height of tghis cylinder is represented at Figure 1. 

\begin{figure}[h!]
\includegraphics[width = 1\textwidth]{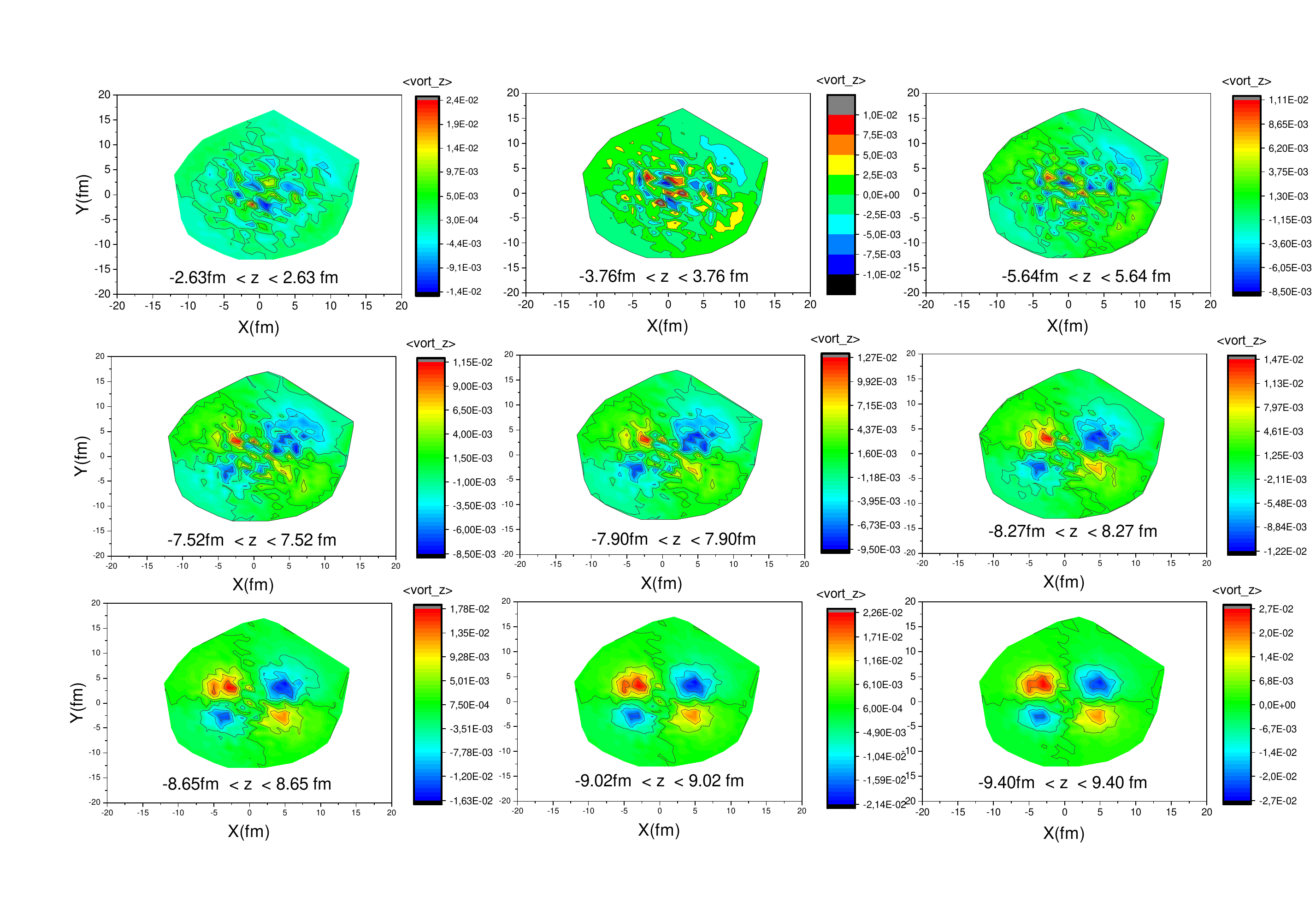}
\caption{Quadrupole structure of average longitudinal vorticity for different heights of the cylinders in $z$ direction (see text).}
\end{figure}

Later more detailed calculations were performed \cite{Baznat:2015eca},
including the structure of emergent vortex sheets, as well as 
spatial and temporal dependence of strange chemical
potential which is also the ingredient of anomalous approach to
polarization. 

\section{Anomalous mechanism of hyperon polarization}

Anomalous mechanism of polarization makes this effect qualitatively
similar to the local violation \cite{Fukushima:2008xe} of discrete
symmetries in strongly interacting QCD matter. The most well known
here is the Chiral Magnetic Effect (CME) which uses the
(C)P-violating (electro)magnetic field emerging in heavy ion
collisions in order to probe the (C)P-odd effects in QCD matter.

The polarization, in turn, is similar to Chiral Vortical Effect
(CVE)\cite{Kharzeev:2007tn} due to coupling to P-odd medium
vorticity leading to the induced electromagnetic and all
conserved-charge currents \cite{Rogachevsky:2010ys}, in particular
the baryonic one.

Recently the pioneering preliminary results on global polarization
of $\Lambda$ and $\bar \Lambda$ hyperons in $Au-Au$ collisions in
Beam Energy Scan at RHIC were released \cite{Lisa} and the
qualitative tendency of polarization decrease with energy in
agreement with the prediction \cite{Rogachevsky:2010ys} was
revealed.  The recent theoretical analysis \cite{Sorin:2016smp}
suggested that decrease of polarization with energy is related to
the suppression of Axial Magnetic effect contribution in strongly
correlated QCD matter found in lattice simulations.

Indeed, the chiral vorticity coefficient describing the axial
vortical effect
\begin{equation}
\label{cv} c_V=\frac{\mu_s^2+\mu_A^2}{2 \pi^2}+\frac{T^2}{6},\quad
\end{equation}
contains temperature dependent term related to gravitational anomaly
\cite{Landsteiner:2011iq}, and naively it can be quite substantial
and increase with energy. However, lattice simulations
\cite{Braguta:2016pwq} lead to the zero result in the confined phase
and to the suppression by one order of magnitude at high
temperatures. Neglecting axial chemical potential, the coefficient
$c_V$ takes the form
\begin{equation}
\label{cvl} c_V=\frac{\mu_s^2}{2 \pi^2}+k \frac{T^2}{6},\quad
\end{equation}

.

As soon as for free fermion gas the $T^2/6$ term is recovered
\cite{Buividovich:2013jba} for large lattice volume at fixed
temperature, the above-mentioned suppression should be attributed to
the correlation effects. It was suggested\cite{Sorin:2016smp}, that
the accurate measurements of polarization energy dependence may
serve a sensitive probe of strongly correlated QCD matter. In the
current paper we perform numerical simulations to implement this
suggestion.

The polarization is related \cite{Baznat:2013zx,Sorin:2016smp}  to the strange axial charge 
\begin{equation}
\label{q5} 
Q_5^s=N_c \int d^3 x \,c_V \gamma^2 \epsilon^{i j k}v_{i}
\partial_{j}v_ k,
\end{equation}
and as a result the quark and hadronic observables are related, that is of special importance in the confined phase. Another approach to confined phase is provided by consideration \cite{Teryaev:2017nro}
of the vortices in pionic superfluid, whose cores are associated with polarized baryons

\section{Numerical simulations of axial anomaly contributions to (anti)hyperon polarization}

We performed the numerical simulations in QGSM model
\cite{toneev83,toneev90,amelin91}. We decomposed the space-time to
the cells, allowing to define velocity and vorticity in the kinetic
model, as described in detail in \cite{Baznat:2013zx}. To define the
strange chemical potential (assuming that $\Lambda$ polarization is
carried by strange quark) we used the matching procedure
\cite{Baznat:2015eca} of distribution functions to its (local)
equilibrium values. In this paper, we also determine in this way the
values of temperature. In general, let us stress that we realized in
our particular case the relation between kinetics, hydrodynamics and
thermodynamics.

We first neglect the gravitational anomaly contribution and start by
considering the energy dependence of polarization (described in
detail in \cite{Sorin:2016smp}) for three values of impact
parameter. The results are presented at Figure 2. The curves correspond
to $b=8.0 fm, 6.4 fm, 4.8 fm$.

\begin{figure}[h!]
\includegraphics
[width=0.8\textwidth]{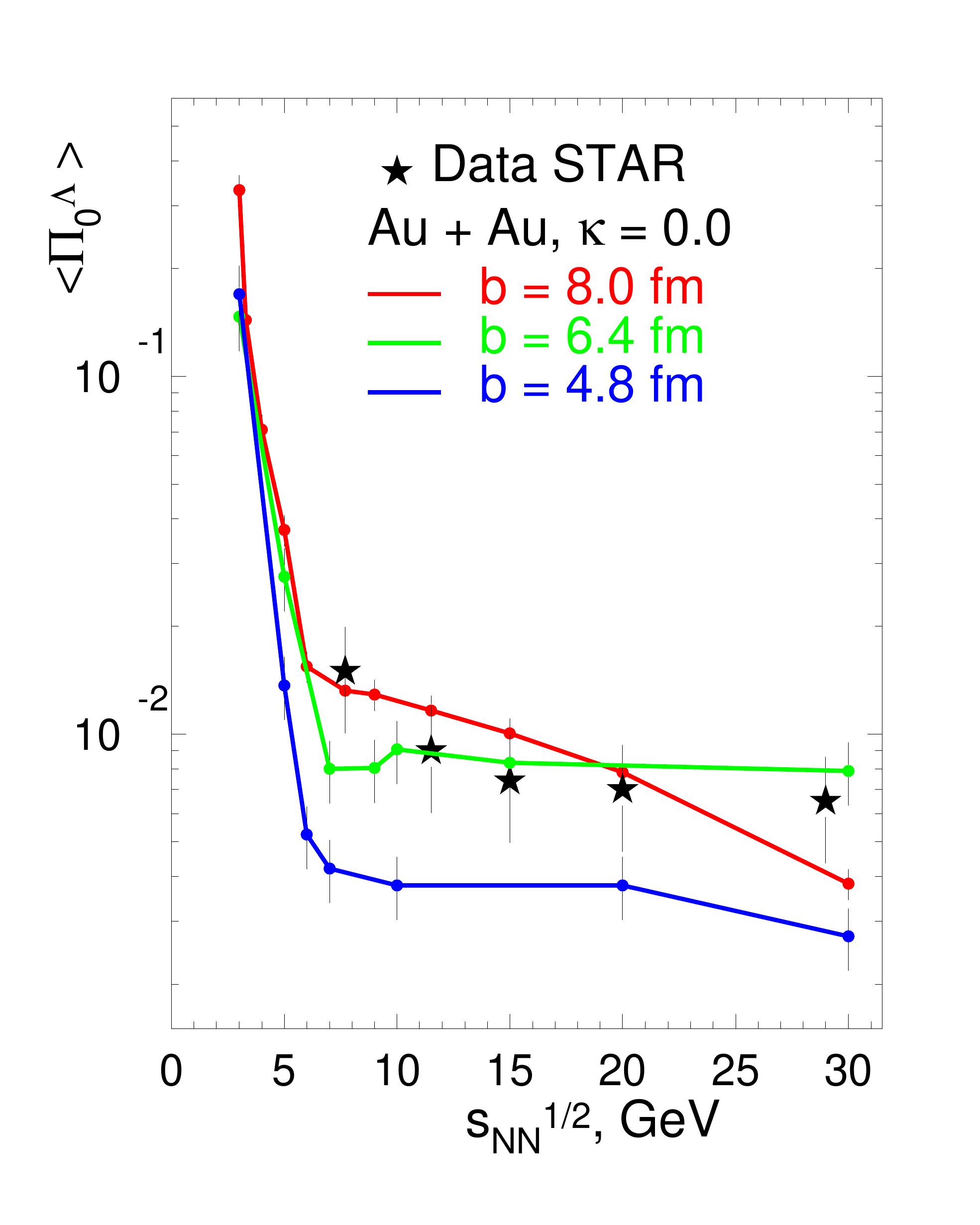}
\caption{Energy dependence of polarization for different values of
impact parameter.}
\end{figure}

We continue by the inclusion of contribution related to
gravitational anomaly, which is the central issue of this paper. The
results are presented at Figure 3. We consider as a starting point the
original value of anomaly coefficient\cite{Landsteiner:2011iq}
$T^2/6$ which is reproduced for large lattice volume at fixed
temperature\cite {Buividovich:2013jba}.
We present the curves following from the coefficients suppressed by
factor $k$ (\ref{cvl})  resulting from the lattice calculations
\cite{Braguta:2016pwq}. We compare values of $k=1$ with $k=0, 1/15,
1/10$. As one can see, the lattice-supported value $1/15$ is most
close to the behavior of preliminary data which may be considered as
a signal of strongly correlated matter formation. The closeness of
$k=0$ curve to the experimental points may be related to the
contribution of confinement phase, where lattice calculations
\cite{Braguta:2016pwq} lead to zero temperature-dependent effect. At
the same time, already $k=1/10$ leads to the curve growing with
energy, while $k=1$ leads to extremely strong growth.

\begin{figure}[h!]
\includegraphics[width=0.8\textwidth]{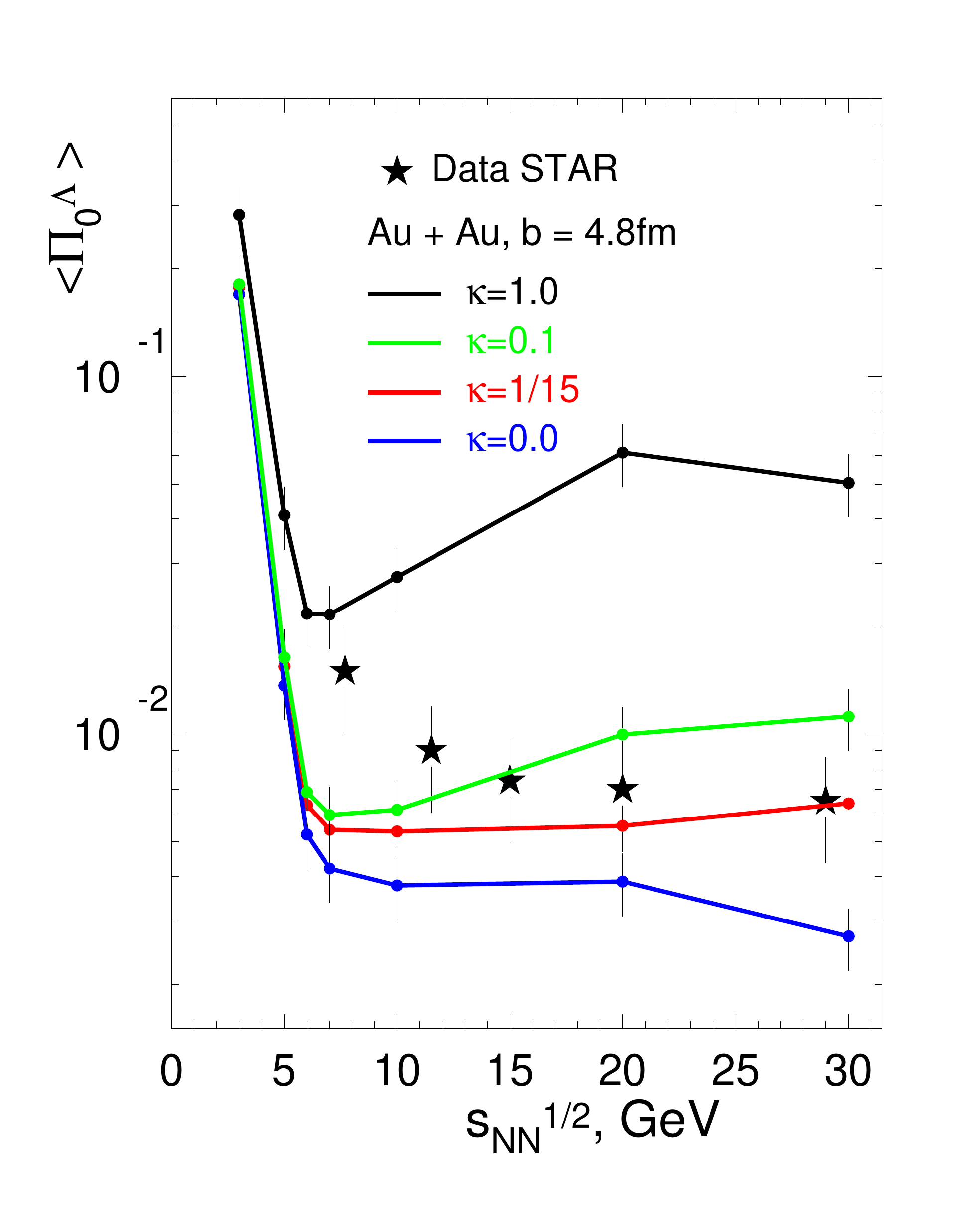}
 \hspace{-20mm}
\caption{Energy dependence of polarization for different values of
gravitational anomaly contribution.}
\end{figure}

The $\bar \Lambda$ polarization is emerging due to the polarization of $\bar s-$quarks, which has the same sign, as the axial current and charge are C-even operators. The magnitude of the $\bar \Lambda$ 
is larger as the same axial charge is distributed between the polarizations of the smaller number of particles \cite{Sorin:2016smp}. It is mandatory to take into account the contribution of $K^*$ mesons. 
In the case of $\Lambda$ the $K^{*-}, \bar K^0$ mesons contain two sea(anti)quarks and does not change the polarization significantly. At the same time, for $\bar \Lambda$ the relevant $K^{*+}, K^0$
mesons are more numerous and make the excess of $\bar \Lambda$ polarization less pronounced. 

Note that this excess is larger for smaller energies, where suppression of $\bar \Lambda$ is larger. 
This differs from the (C-odd) effect of magnetic field, which is increasing with energy, although more detailed studies taking into account the finite time of magnetic field existence  are required. 

The quantitative analysis of these effects, taking into account the gravitational anomaly contribution, is presented at Figure 4. 
\begin{figure}[h!]
\includegraphics[width=0.8\textwidth]{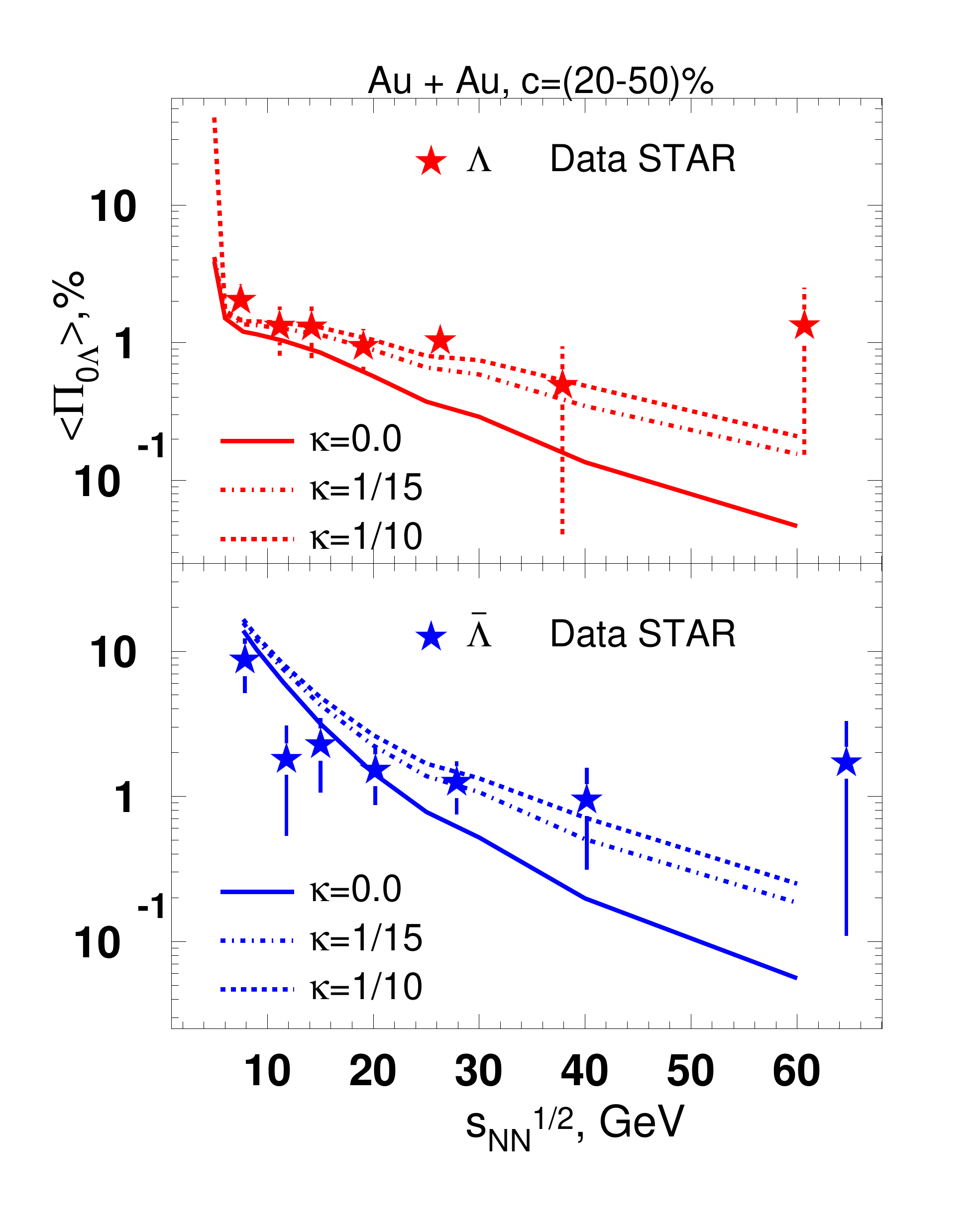}
\caption{Comparison of $\Lambda$ and $\bar \Lambda$ polarizations for different values of
gravitational anomaly contribution.}
\end{figure}
The result is in reasonable agreement with STAR data, although further analysis 
is required.

\section{Conclusions and Outlook}


We numerically studied the generation of polarization by the
anomalous mechanism (Axial Vortical Effect) and compared it with the
observed data.

First we neglected the gravitational anomaly related temperature
dependent contribution when the decrease of chemical potential with
energy leads, in turn, to the decrease of polarization. We considered
this effect for three impact parameters.

We also  included the contribution related to gravitational anomaly
proportional to $T^2$ and studied its possible suppression in
strongly correlated matter. We found that the preliminary data are
in accordance with suppression effect found on the lattice.

We also considered the polarization of $\bar \Lambda$ hyperons, taking into account 
the contribution of $K^*$ mesons. We found that the $\bar \Lambda$ polarization is larger than 
that of  $\Lambda$ and is growing at smaller energies.

The further more accurate measurements of $\Lambda$ polarization
should provide the additional check of gravitational anomaly related
contribution.

%
The useful discussions with F. Becattini, V. Braguta, S. Voloshin and V.Zakharov are gratefully acknowledged.
This work was supported in part by the Russian Foundation for Basic
Research, Grant No. 17-02-01108.


\end{document}